\documentclass[12pt]{article}
\pdfoutput=1 
\usepackage{sbc-template}
\usepackage{graphicx,url}
\usepackage[utf8]{inputenc}
\usepackage[brazil]{babel}
     
\title{Student similarity network clustering - Does the time it takes for an answer to be selected follow a power-law?}

\author{Filipe S. P. Prates\inst{1}}

\address{Universidade Federal do Rio de Janeiro
  (UFRJ)\\  
  Caixa Postal 68511 CEP: 21941-972  Rio de Janeiro - RJ - Brasil
  \email{filipeprates@poli.ufrj.br}
}

\begin{document} 

\maketitle

\begin{abstract}
  This article uses a dataset of answers to questions to generate student similarity networks. Two similarity functions to determine the weights between each pair of students are used, one that assumes a power-law distribution of answers \textit{response times}, and one that does not. The resulting networks are then clustered using different community finding algorithms and their resulting modularity is compared.
\end{abstract}
     
\begin{resumo} 
    Este artigo utiliza uma base de dados de respostas a questões escolares para gerar uma rede de similaridade entre os estudantes. Nele é utilizando duas funções de similaridade distintas para determinar o peso da aresta existente entre cada par de alunos, uma que assume uma distribuição em lei de potência dos \textit{tempos de resposta}, e uma que não. As redes resultantes serão então sujeitas à algoritmos de agrupamento e as modularidades resultantes comparadas.
    \end{resumo}

\section{Introduction}

Using data of student answers to quizzes we measure and understand how the different students behaviors can be clustered. Each answer in the dataset is related to a student and a question, and it carries the information about whether it was correct (\textit{value}) and the time the student took to answer it (\textit{response time}). 
In this article we are going to create student similarity networks, where students edges are weighted by the similarity of their answers, then use different clustering algorithms on the resulting networks and compare their results.
I propose that the distribution of the \textit{response time} of the answers follow a power-law, to test if that is the case two similarity metrics are created, one that assumes a power-law distribution, and one that does not. We then compare the modularity of the resulting clusterings of the student similarity network. If the \textit{response time} follows a power-law distribution, we expect the algorithms to find better (higher modularity) partitions.

\section{Methodology}

\begin{figure}
    \centering
    \includegraphics[width=1.0\linewidth]{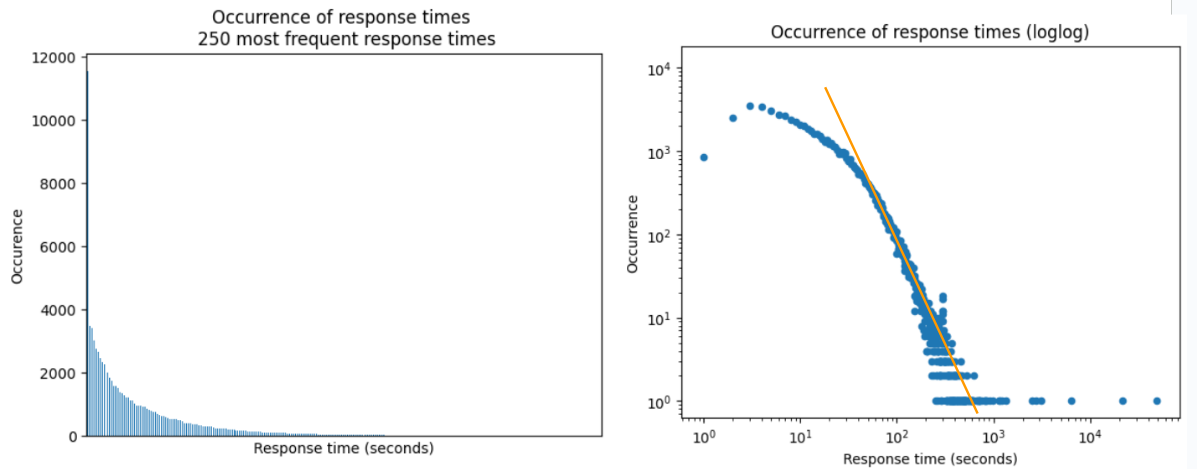}
    \caption{Frequency analysis of \textit{response time} of answers available in the dataset. Observed linear behaviour in log-log plot hinting to a possible power-law distribution. }
    \label{fig:occRT}
\end{figure}

Observing the 88000 answers made available by Jovens Gênios Provedor de Conteúdo LTDA, we can see an interesting distribution of \textit{response times}, where 0.0 seconds is the most frequent indicating a random guess by the students, followed by very frequent low \textit{response time} values, and less frequent but present high \textit{response time} values. (Figure~\ref{fig:occRT})

Observing all answers that cover a broad range of knowledge areas, we can observe a linear behaviour in the occurence of the response times in a log-log plot, which prompted the experiment to take this power law into consideration when developing the similarity function.

\begin{figure}
    \centering
    \includegraphics[width=0.5\linewidth]{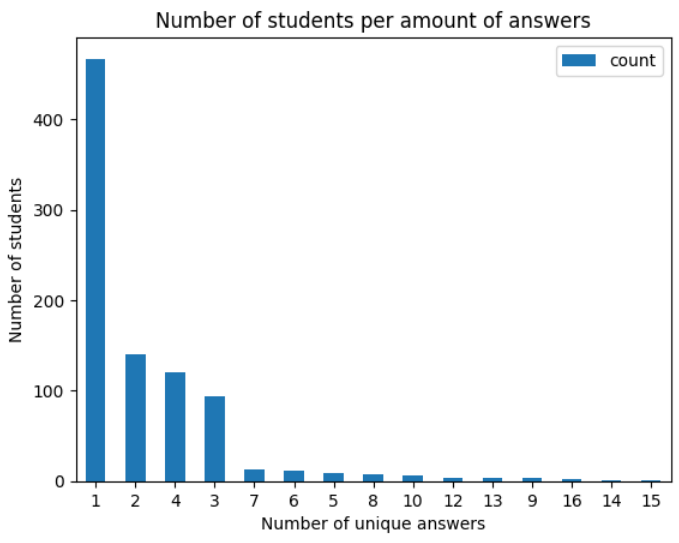}
    \caption{Over half the students (467/882) have only one answer in the dataset, the vast majority answered at most four times (822/882), a small but significant number have repeated answers to a same question at least once (60/882).}
    \label{fig:stuAnsw}
\end{figure}

For this study, 2000 answers from a single knowledge area where chosen. These answers pretain to 882 unique students answering 4 unique questions, in the span of one year. Some students have answered just one of the questions, while others multiple and even repeated ones. (Figure~\ref{fig:stuAnsw})

\subsection{Similarity functions}

To measure the similarity between two students we are going to compare each of their answers in pairs. For each pair, if both answers are related to the same question, and have the same \textit{value} (a boolean that indicates if the answer was correct or not), then we compare their \textit{response times} and add the resulting value to the weight of their edge.

$$ \gamma(r_i,r_j) = (r_i.value \oplus r_j.value)! $$
$$ \alpha(r_i,r_j) = (r_i.question\_id  \oplus r_j.question\_id)! $$
$$ \beta(r_i,r_j) =  Sim_{tr}(r_i, r_j) $$
$$ Sim_{a}(a_i,a_j) = \sum_{r_i \in a_i.answers}\sum_{r_j \in a_j.answers}\gamma(r_i,r_j) \alpha(r_i,r_j)\beta(r_i,r_j) $$
The similarity between students \(a_i\) and \(a_j\) is determined by the similarity of their answers \(r_i\) and \(r_j\).
For this article we are going to compare two answer \textit{response time} similarity functions:
$$  Sim_{rt_0}(r_i,r_j) = 1 - \frac{|r_j.response\_time - r_i.response\_time|} 
{maxT} $$
and 
$$  Sim_{rt}(r_i,r_j) = 1 - \frac{|log(r_j.response\_time + 1) - log(r_i.response\_time + 1)|}{log (maxT + 1)}$$. Where maxT is the maximum \textit{response time} found in the dataset.

While the first one assumes that the similarity between two number depends solely on the absolute difference of their values, the second one takes into account the magnitude of each value.
If it is true that the \textit{response times} follow a power-law distribution, we expect the second function to better determine the distance between two answers, as two low values a certain absolute difference apart should be less similar than two high values with the same absolute difference, similarly to their occurrences in the dataset.

The distribution of \textit{response times} is not linear from \([0, maxT]\), and so we want our similarity metric \(Sim_{rt}(r_i,r_j)\) to reflect that.
The resulting strength distribution of the student nodes (sum of weights associated with each student) is given by (Figure~\ref{fig:stuAnsw})
\begin{figure}
    \centering
    \includegraphics[width=0.5\linewidth]{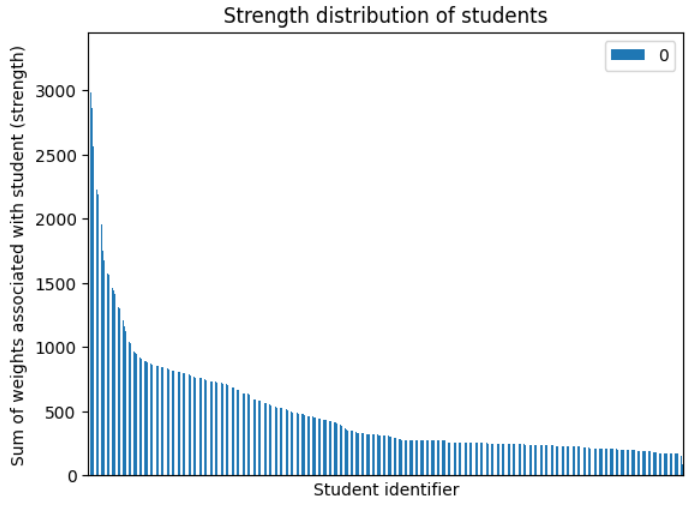}
    \caption{Distribution of strength (sum of weights of their edges) of student nodes in student similarity network using \(Sim_a(a_i,a_j)\).}
    \label{fig:strStu}
\end{figure}

\subsection{Algorithms and Graphs}

\begin{figure}
    \centering
    \includegraphics[width=0.5\linewidth]{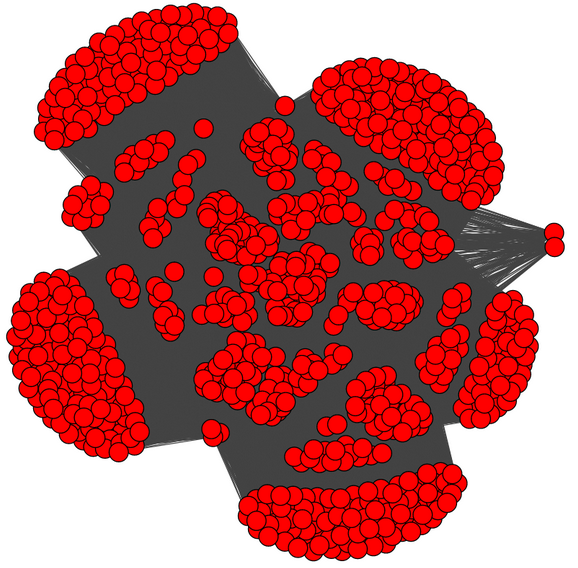}
    \caption{ Python igraph library default plot of student similarity network for 2000 answer dataset. }
    \label{fig:g}
\end{figure}

To generate the student similarity network, each student found in the dataset was represented as a node. Then for each pair of students we calculated the similarity between them (\(Sim_a(a_i,a_j)\)) and the resulting value determined the weight of their respective edge.
The resulting network has 882 nodes and 164219 edges (pairs of students which had at least one answer to the same question with the same \textit{value}). (Figure~\ref{fig:g})

To cluster the resulting graph we are going to use the following algorithms:
\begin{itemize}
    \item Infomap \cite{rosvall2008maps}\cite{rosvall2009map}
    \item Fast and Greedy \cite{clauset2004finding}
    \item Label Propagation \cite{raghavan2007near}
    \item Leading Eigenvector \cite{newman2006finding}
    \item Multilevel \cite{blondel2008fast}
    \item Walktrap \cite{pons2005computing}\cite{newman2005power}
\end{itemize}

Both the student similarity network created with \(Sim_{rt0}(r_i,r_j)\) and \(Sim_{rt}(r_i,r_j)\) will be subjected to the algorithms with the same parameters and the resulting modularity \cite{newman2006modularity} and cluster numbers compared.

\section{Results}

\begin{figure}
    \centering
    \includegraphics[width=1\linewidth]{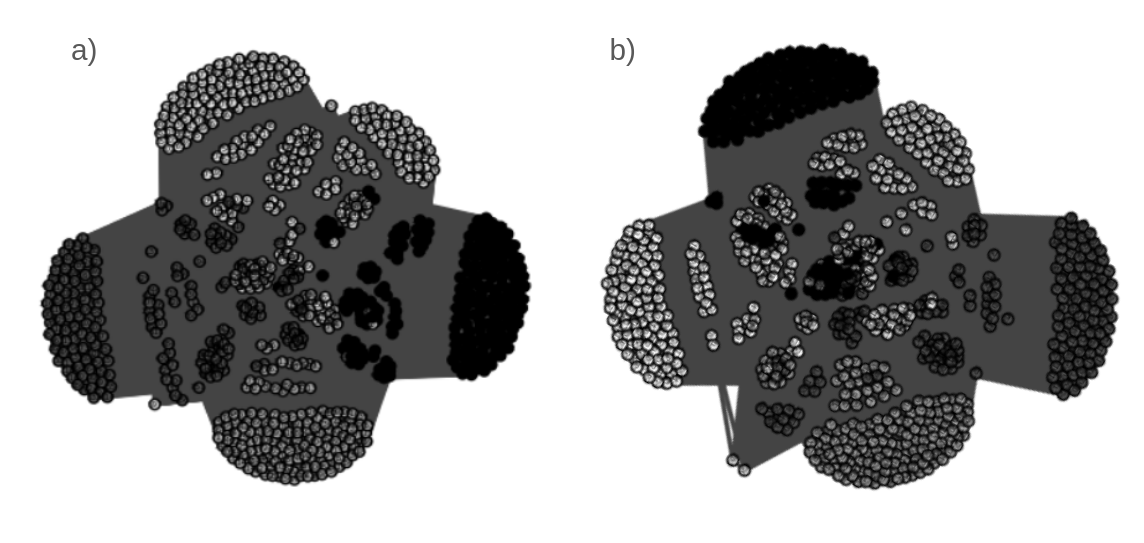}
    \caption{Student similarity network clustered by Multilevel algorithm. On a)  \(Sim_{rt}(r_i,r_j)\), which takes into account the proposed power-law distribution of answers' \textit{response times}, we have 4 clusters and a modularity of 0.278. On b) \(Sim_{rt0}(r_i,r_j)\), where we only take into account the absolute difference of \textit{response times}, we have 5 clusters and a modularity of 0.261.}
    \label{fig:comparingML}
\end{figure}

The following results were obtained:
\begin{center}
\begin{tabular}{ |c|c|c| } 
\hline
Algorithm & Modularity using \(Sim_{rt_0}\) & Modularity using \(Sim_{rt}\) \\
\hline
InfoMap & 0.0 & 0.0 \\
Fast and Greedy & 0.220 & 0.241 \\ 
Label Propagation & 0.0 & 0.0 \\ 
Leading Eigenvector & 0.253 & 0.258 \\ 
Multilevel & 0.261 & 0.278 \\
Walktrap & 0.196 & 0.187 \\
\hline
\end{tabular}
\end{center}
\begin{center}
\begin{tabular}{ |c|c|c| } 
\hline
Algorithm & Clusters using \(Sim_{rt_0}\) & Clusters using \(Sim_{rt}\) \\
\hline
InfoMap & 1 & 1 \\
Fast and Greedy & 3 & 4 \\ 
Label Propagation & 1 & 1 \\ 
Leading Eigenvector & 4 & 4 \\ 
Multilevel & 5 & 4 \\
Walktrap & 4 & 3 \\
\hline
\end{tabular}

\end{center}
We achieved an average of 2.24\% higher modularity in the clusterings across all algorithms when using the \(Sim_{rt}(r_i,r_j)\), the similarity of \textit{response times} function that takes into account their observed power-law distribution. Excluding the Label Propagation and Infomap algorithms, which both resulted in a single giant cluster regardless of the similarity function (modularity of zero), we had a 3.36\% increase in modularity when applying the \(Sim_{rt}(r_i,r_j)\) function. (Figure~\ref{fig:comparingML})

Selecting only the algorithms that improved modularity when utilizing the metric, Fast and Greedy, Leading Eigenvector, and Multilevel, we had a relative increase in modularity of 5.84\%.  But there was one algorithm that performed significantly worse when using \(Sim_{rt}(r_i,r_j)\), the Walktrap, which had 5.6\% reduction in modularity.

\section{Conclusions}

The high density of the network which reflects very similar students (due mainly to the low amount of unique questions) makes it so the clustering algorithms return a small amount of clusters with relatively low modularity. Even though the algorithms on average achieved better modularity results using \(Sim_{rt}(r_i,r_j)\) (taking into account the proposed power-law distribution of \textit{response times}) the result of the Walktrap algorithm showed that the experiment was not conclusive to determine if \textit{response times} really follow a power-law and if the \(Sim_{rt}(r_i,r_j)\) metric was enough to encode this distribution information in a way to help the algorithms better determine the similarity between students given their answers.

It's worth to note that the way the similarity of students (\(Sim_a(a_i,a_j)\)) function is structured, repeated answers to the same question between pairs of students result in a much higher weight between them compared to when they answer unique questions everytime. More specifically, for every pair of students, if each has \(k\) answers to a question \(Q\) (and assuming same \textit{value} and \textit{response time} for all of them), the weight of the edge between them due to the question Q will be \(k^2\). This results in easily clustered outlier students (possible improper use of the quiz interface) which have much higher weight in the edge between them compared to other student pairs.

\section{Observations}

The 88000 dataset of answers utilized to create (Figure~\ref{fig:occRT}) were randomly selected from Jovens Gênios database, but the 2000 answer dataset used to create (Figure~\ref{fig:stuAnsw}) and the resulting student similarity network were selected as the first 2000 answers in the current year in a specific Course, following the question identifier order. This resulted in the small amount of unique questions present in the data. It's important to note that the first dataset lacks information identifying the students associated with the questions. Therefore, for a more comprehensive investigation of the resulting network, a new dataset including student identifiers relating to a broader range of questions would be necessary.

\section{Future studies}

A larger study with more answers that span a broader array of subjects, taking into account the abilities and topics that each of the questions develops in the similarity of students functions would result in a more complex and deeper understanding of the students behaviour.

\bibliographystyle{sbc}
\bibliography{sbc-template}

\end{document}